\documentclass[epsf,twocolumn,preprintnumbers]{revtex4}
\usepackage{graphics}
\usepackage{graphicx}% Include figure files
\usepackage{dcolumn} % Align table columns on decimal point
\usepackage{bm}
\usepackage{color}
\usepackage{epsfig}
\newcommand{\vop}{V$_2$OPO$_4$}

\begin{document}
\title{Charge and Orbital Orderings, and Frustration in Quasi-one-dimensional Ferrimagnetic Insulator 
 $\beta$-V$_2$O(PO$_4$) 
}
\author{Seo-Jin Kim$^1$}
\author{Kwan-Woo Lee$^{1,2}$}
\email{mckwan@korea.ac.kr}
\affiliation{
 $^1$Department of Applied Physics, Graduate School, Korea University, Sejong 30019, Korea\\
 $^2$Division of Display and Semiconductor Physics, Korea University, Sejong 30019, Korea
}
\date{\today}
\begin{abstract}
Using {\it ab initio} calculations based on the correlated band theory,
we have investigated 
%the electronic and magnetic structures of 
the quasi-one-dimensional chain system $\beta$-V$_2$O(PO$_4$), 
showing both charge and spin orderings. 
%at T$_{CO}\approx 600$ K and T$_{SO}\approx 128$ K respectively.
Even in the uncorrelated region,
the pure transition from the tetragonal to the monoclinic structure
leads to a sizable charge difference between the two types of V ions,
regardless of magnetic orders.
In the ferrimagnetic phase, 
inclusion of the on-site Coulomb repulsion $U$ leads to a full orbital-polarization 
of V1 ($t_{2g}^{3\uparrow}$, $S=\frac{3}{2}$) and V2 ($a_{1g}^{1\downarrow}e_g^{\prime{1\downarrow}}$, $S=1$) 
above $U^c_{eff}\approx3.5$ eV, leading to local spin moments of 2.30 and --1.54 $\mu_B$, respectively,
with small orbital moments of several hundredth $\mu_B$.
So, the net moment is nearly 1 $\mu_B$ per formula unit, which is about 2--3 times larger than the experimental value.
%However, the orbital order in V2 ion leads to a small orbital moment less than tenth $\mu_B$. 
Our results show significant variations, strongly depending on the strength of $U_{eff}$, 
in energy differences between various magnetic states as well as a small magnetic anisotropy.
These results suggest that the substantial difference between the calculated and experimental moments 
is attributed to quantum fluctuation of the pyrochlore-like weakly linked V$_4$ tetrahedral structure.
Our findings are expected to provide a good platform to investigate the interplay 
among the charge-, spin-, and lattice-degrees of freedom, and geometrical frustration.
\end{abstract}
\maketitle

\section{Introduction}
In condensed matters, the interplay among the charge-, spin-, orbital-, and lattice-degrees of freedom
leads to abundant interesting phenomena. 
Over the past several decades, low-dimensional spin systems have been intensively investigated
due to their exotic phenomena such as spin-Peierls (or dimers),\cite{gros03}
spinon confinement,\cite{wang15} and even high T$_c$ superconductivity.\cite{rice96}
%When the low dimensionality coexists with frustrated structures,
%the competition between the low dimensionality and frustration leads 
%to various exotic quantum magnetic states.\cite{hida08,star15} 
In particular, quasi-one-dimensional (1D) chains with (weakly) frustrated structures 
have been attracted significant interest 
owing to various characteristic quantum magnetic states driven by the competition between the low dimensionality 
and frustration.\cite{hida08,star15}
Most studies have been carried out on 1D  chains with weakly linked triangular lattices,
{\it e.g.}, Ca$_3$Co$_2$O$_6$,\cite{hardy03,hardy04} $\beta-$TeVO$_4$,\cite{tvo11,tvo16} 
and Sr$_2$Rh$_4$O$_{12}$-family.\cite{parkin07}
Another example is the trans-1,4-cyclohexanedicarboxylate of alternating Fe$^{2+}$ and Fe$^{3+}$ 
with a 1D chain and spin ladder structure.\cite{ladder}

Many vanadium oxides exhibit metal-insulator transition (MIT), 
charge ordering (CO), and even orbital ordering, but their mechanisms have been debated
over the past several decades.
For example, the binary and perovskite-type ternary vanadium oxides 
such as VO$_2$ (with V$^{4+}$ $d^1$) and V$_2$O$_3$ (with V$^{3+}$ $d^2$)\cite{cryfield,v2o3_06,v2o3_08,yu99,vo2_dmft,v2o3_18} 
have been intensively investigated
owing to their peculiar MITs and possible applications to information technology and, 
electronic and photonic devices.\cite{mrs17}
In the quasi-1D paramagnetic VO$_2$, MIT is induced by the dimeralization of V ions.\cite{vo2_dmft}
On the other hand, for the antiferromagnetic monoclinic V$_2$O$_3$
it is not clear whether MIT accompanies orbital-ordering.\cite{cryfield,v2o3_06,v2o3_08,v2o3_18}
In addition, V$_4$O$_7$ consisting of two independent cation chains
exhibits an unconventional MIT involved in dimeralized spin-singlet V$^{4+}$ ions and magnetically ordered V$^{3+}$.\cite{botana11}

In this paper, we provide a promising candidate 
of a distinct low-dimensional and frustrated system showing
{\it a quasi-1D ferrimagnetic Heisenberg spin chain structure
with a weak pyrochlore-like tetrahedral link} of vanadium ions in the $\beta$-\vop~ vanadium phosphate.
In the formal charge concept of (PO$_4$)$^{3-}$ and O$^{2-}$, 
V$_2$O(PO$_4$) is a mixed valent system with two V$^{2.5+}$ ions in average.  
As will be addressed below, 
this system shows charge and spin orderings intertwined with the lattice degree of freedom.
Therefore, below the spin ordering temperature, there are chains of alternating 
V$^{2+}$ ($d^3$, $S=\frac{3}{2}$) and V$^{3+}$ ($d^2$, $S=1$) ions.
The chain structure exists more rarely than in a V$^{4+}$ system.

\vop~ was initially reported to be tetragonal at room temperature.\cite{glaum}
Based on the assumption of a ferromagnetic (FM) state in the tetragonal structure,
through theoretical investigations,
Jin {\it et al.} proposed that this system is a magnetic Weyl semimetal.\cite{y.jin}
However, recently,  Pachoud {\it et al.}\cite{attfield} revealed
a tetragonal-to-monoclinic structure transition at $T_{s}\approx$600 K,
leading to two distinct V sites.
At $T_{s}$, a transition of negative to positive thermal expansion simultaneously occurs.
Through a resistivity measurement demonstrating the semiconducting behaviors in both phases 
and analysis of the temperature-dependent bond valence sums,
they concluded that CO occurred in the 1D V-V chain at $T_{s}(=T_{CO})$.
This suggested that the very rare negative thermal expansion in the high-temperature (tetragonal) phase
was driven by loss of CO.
%The oxidation state of V ions is $+2.5$, in average, since (PO$_4$)$^{3-}$ in the formal charge concept.
In the low-temperature (monoclinic) phase, a ferrimagnetic (FI) spin ordering is observed 
at the Curie temperature of $T_C=164$ K.
Isothermal magnetization measurements indicated the FI order between nearest neighbor (NN) V ions,
which was confirmed by a neutron powder diffraction measurement.
Xing {\it et al.}\cite{j.xing} confirmed most of these results with a single crystalline sample,
but observed a lower $T_C=128$ K than that of the polycrystalline sample.
The magnetic susceptibility above $T_C$ provides a high Curie-Weiss temperature of $\Theta_{CW}=-900$ K
and effective moment of 3.7 $\mu_B$ per V ion.
The effective moment is closer to the value of the V$^{2+}$ ion than to that of the V$^{2.5+}$ ion,
implying a complex magnetic interaction.
The total moment, incompletely saturated even at 6 T, is 0.27 -- 0.31 $\mu_B$/V 
with a remnant moment of 0.22 -- 0.25 $\mu_B$,
comparable to the value of 0.6 $\mu_B$/V obtained by Pachoud {\it et al}.\cite{attfield} 
The magnitudes of the V local moments are approximately 1.2 -- 1.5 $\mu_B$ for V$^{3+}$ and 1.4 -- 1.8 $\mu_B$ for V$^{2+}$,
which are substantially reduced from the nominal values of $S=1$ and $\frac{3}{2}$, respectively.
From the resistivity data,
the energy gap of 0.48 eV could be estimated.\cite{j.xing} 
%with a large low temperature resistivity of order of $M\Omega\cdot cm$

As mentioned above, this vanadium phosphate shows various peculiar properties.
In addition to the negative thermal expansion in the high T phase, which will not be discussed here,
this system exhibits several interesting properties in the low T phase: (i) 1D mixed spin chain 
with a promising frustrated structure, (ii) CO FI insulator, and
(iii) magnetic moments substantially reduced from the nominal values. 
%of spin 1 and $\frac{3}{2}$.
Through {\it ab initio} calculations, including correlation effects,
we have investigated the electronic and magnetic structures of the low T monoclinic phase,
which has not been previously reported in detail.

\begin{figure}[tbp]
%\vskip 8mm
{\resizebox{8cm}{8cm}{\includegraphics{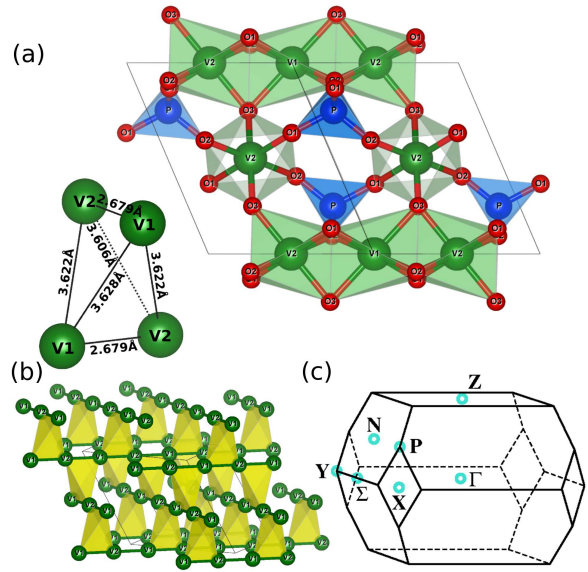}}}
\caption{(a) Side view of the monoclinic structure of V$_2$O(PO$_4$),
consisting of PO$_4$ tetrahedra and VO$_6$ octahedra.
The V ions form a quasi-1D chain.
(b) Network of the corner-sharing V$_4$ tetrahedra. 
As shown in the Inset,
the interchain V--V distances along the vertical direction are approximately 1/3 longer than 
the intrachain value. (For details, see text.)
(c) Conventional Brillouin zone and high symmetry points of the body-centered tetragonal structure,
 as given in Ref. [\onlinecite{y.jin}].
}
\label{str}
\end{figure}

\begin{table}[bt]
\caption{Nearest neighbor interatomic distances in the PO$_4$ tetrahedra and VO$_6$ octahedra  
 (in units of \AA).}
\begin{center}
\begin{tabular}{ll}\hline\hline
  V1$-$O~~  & 2$\times$2.100,~2$\times$2.096,~2$\times$2.084 \\
  V2$-$O~~ & 2$\times$1.989,~2$\times$2.032,~2$\times$2.059 \\
  P$-$O~~ & 4$\times$$\sim$1.53 \\\hline\hline
\end{tabular}
\end{center}
\label{table1}
\end{table}

\section{Crystal structure and calculation methods}
Our calculations were carried out for the monoclinic V$_2$O(PO$_4$) containing two formula units, 
displayed in Fig. \ref{str}(a),
with the experimental lattice parameters of $a=$7.5552 \AA, $b=$7.5979 \AA, $c=$7.2110 \AA,
and the angle $\beta=121.2^\circ$.\cite{attfield}
In the structure (space group: No. 15, $C2/c$), V1 and V2 atoms occupy at $4c~(\frac{1}{4},\frac{1}{4},0)$ and 
$4b~(0,\frac{1}{2},0)$ sites, respectively.
O1 and O2 lie at $4f~(x,y,z)$ sites. O3 and P atoms reside at $4e~(0,y,\frac{1}{4})$ sites.
O1 and O2 are shared by VO$_6$ octahedra and PO$_4$ tetrahedra, 
whereas O3 atoms are vertically connected only to the VO$_6$ octahedra.
We used the internal parameters experimentally obtained at 10 K.\cite{attfield}
As summarized in Table \ref{table1},
the V1-O$_6$ octahedron has an approximately 3\% (in average) longer V-O bond length  
than that in the V2-O$_6$ octahedron.
The octahedron of V2 is more irregular than that of V1,
indicating a substantial trigonal distortion in the V2-O$_6$ octahedra.
These lead to the charge and orbital orderings, as will be shown below.

Basically, the crystal consists of substantially distorted VO$_6$ octahedra 
and nearly regular PO$_4$ tetrahedra, as shown in Fig. \ref{str}(a).
The PO$_4$ tetrahedra, at layers of $b\approx$1/4 and 3/4, are corner- and edge-shared 
with four oxygen ions of the octahedra.
At the $b$=1/2 layer, the V-V chains are stretched in the direction perpendicular to the V-V chains
at the $b$=0 layer.
Here, the chain is on the $ac$-plane.
The face-sharing VO$_6$ octahedra form a quasi-1D chain 
with a notably short intrachain V-V distance of 2.68 \AA.
The interchain V-V ionic distances are 5.35 \AA~(in-plane) and 3.60--3.63 \AA~(interlayer). 
Remarkably, due to the relatively short interlayer interchain V-V distance,
the network of V ions leads to a pyrochlore-like tetrahedral link,
as displayed in Fig. \ref{str}(b).

%In the PO$_4$ tetrahedra, the atomic distance of P-O is 1.53 \AA.
%The various P-O and V-O atomic distances are given in Table \ref{table1}.

The calculations were based on the generalized gradient approximation (GGA)
with the Perdew-Burke-Ernzerhof (PBE) exchange-correlation functional,\cite{gga} 
implemented in the accurate all-electron full-potential {\sc wien\small{2}\normalsize{k}} code.\cite{wien2k}
The correlation effects are treated by the GGA+U approach.\cite{amf}
The strength of the effective on-site Coulomb repulsion $U_{eff}=U-J$ is varied 
in the range of 2 -- 5 eV,\cite{v2o3_06,yu99,vo2_dmft,v2o3_18,botana11}
where $J$ is the Hund's rule coupling.
We also obtained results with  a separate $U$ and $J$ scheme, but both results were very close to each other.
So, only the results obtained from the $U_{eff}$ scheme are presented.
In {\sc wien2k}, the basis size was determined by R$_{mt}$K$_{max}$=7
and augmented-plane-wave sphere radii (1.96 a.u. for V, 1.28 a.u. for O, and 1.61 a.u. for P).
The Brillouin zone was sampled with a $k$-mesh of {\bf $21\times 17\times 21$}.

\begin{figure}[tbp]
%\vskip 8mm
{\resizebox{8cm}{6cm}{\includegraphics{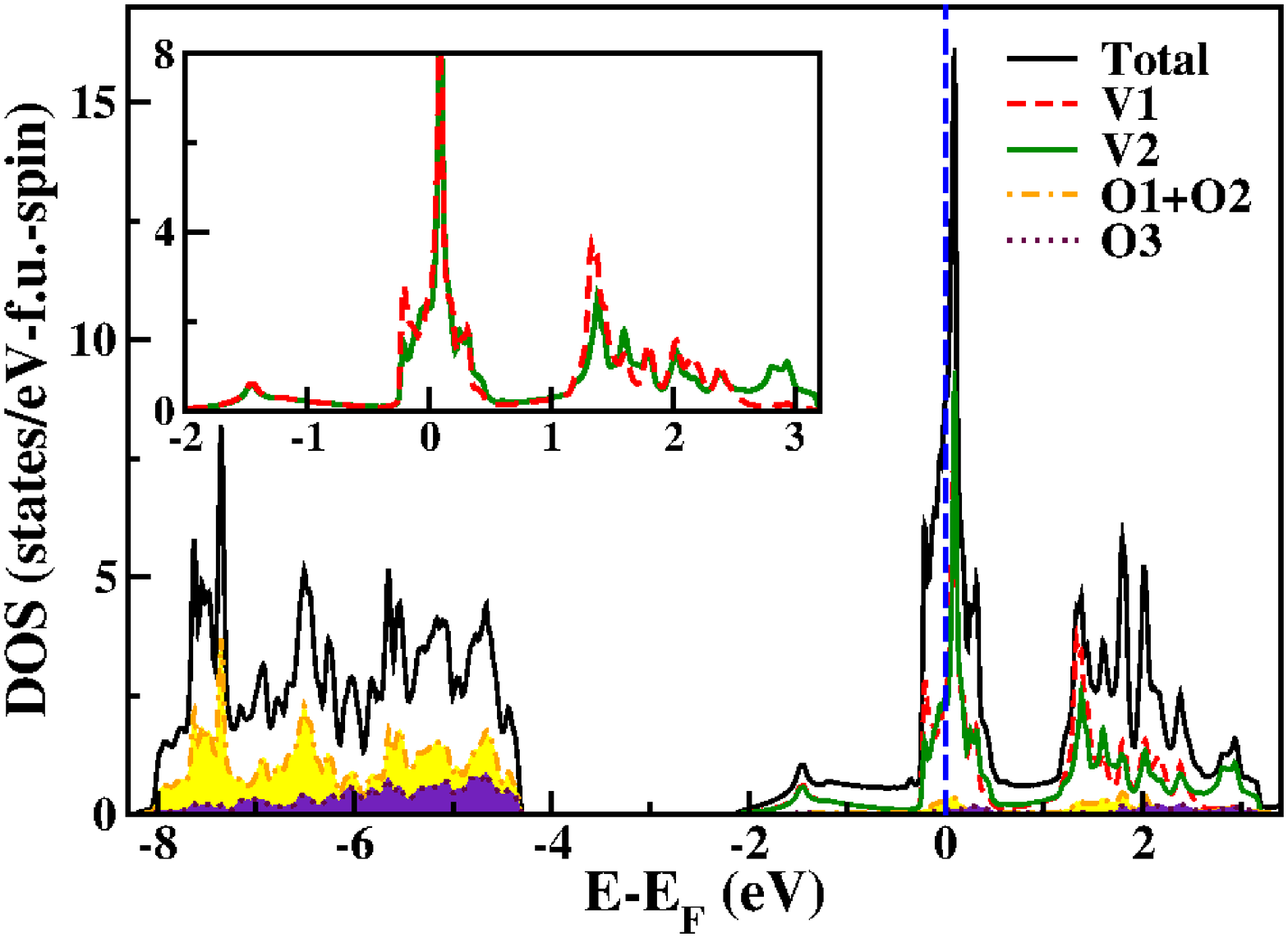}}}
\caption{GGA nonmagnetic total and atom-projected densities of states (DOSs) 
 in the full range in units of formula unit (f.u.). 
 The small DOS of P ion is not shown here.
 The Fermi energy $E_F$, which is denoted by the vertical dashed line set at zero,
lies just below the sharp peak, 
leading to a large total DOS $N(E_F)=$8.72 states per eV per f.u. per spin at $E_F$.
 Inset: Blowup V1 and V2 atom-projected DOSs around $E_F$.
 Both are similar, but some clear distinctions are visible above $E_F$,
 indicating the CO. 
}
\label{pmdos}
\end{figure}

\section{Underlying Electronic Structure}
Before considering the complications of spin order and correlation effects,
we consider the nonmagnetic state within the GGA level to understand 
the basic underlying features of the electronic structure.

Figure \ref{pmdos} shows the nonmagnetic total and atom-resolved densities of states (DOSs),
indicating a (V$_2$)$^{5+}$ configuration, as expected from the formal charge.
The V $d$ orbitals in the range of --2 to 3.2 eV 
are separated with a large hybridization gap of 2.2 eV from the O $p$ orbitals, 
which are spread in the range of --8 to --4.2 eV.
%The $d$ orbitals existing --2 eV to 3.2 eV leads to a sharp peak around the Fermi level $E_F$.
The VO$_6$ octahedral structure leads 
to a crystal field splitting of triplet $t_{2g}$ and doublet $e_{g}$ manifolds.
The crystal field splitting measured from the center of each manifold 
is about 2.5 eV.
In this system, the triplet $t_{2g}$ manifold splits into the doublet $e_g^\prime$ (or $e_g^\pi$) 
and singlet $a_{1g}$ orbitals due to the sizable trigonal distortion in the octahedra.
Setting the $z$-axis along the V-V chain, 
the $a_{1g}$ orbital has a $d_{z^2}$-like shape,
thus leading to a strong V-V direct interaction along the chain.
The $e_g^\prime$ orbitals can be represented by\cite{cryfield} 
\begin{eqnarray}
e_{g,1}^\prime&=&\frac{1}{\sqrt3}(d_{zx}-{\sqrt2}d_{x^2-y^2}),\nonumber\\
e_{g,2}^\prime&=&\frac{1}{\sqrt3}(d_{yz}+{\sqrt2}d_{xy}). 
\label{eqn}
\end{eqnarray}
A small monoclinic distortion, as that in this system, leads to a small splitting of the $e_g^\prime$ orbitals.

\begin{figure}[tbp]
%\vskip 8mm
{\resizebox{8cm}{6cm}{\includegraphics{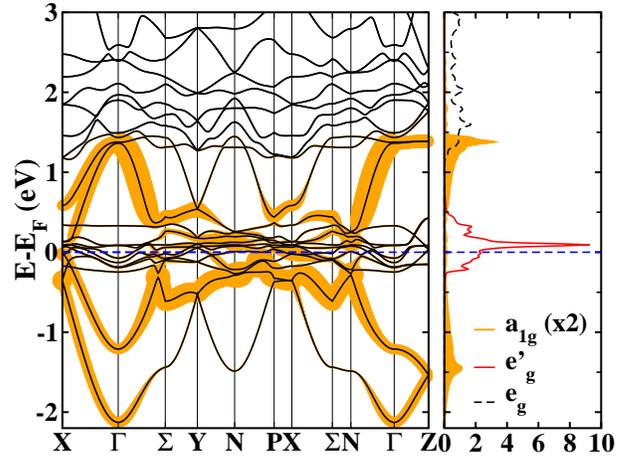}}}
\caption{Left: GGA nonmagnetic band structure enlarged in the range of --2 eV to 3 eV
containing only V $3d$ orbitals.
The character of the $a_{1g}$ orbitals is highlighted.
Right: Orbital-projected DOSs of V2 $3d$ orbitals, in units of states per eV per atom.
For a better visualization, that of the $a_{1g}$ is enhanced by a factor of two.
The V1 correspondence is similar, and thus is not shown here.
The high symmetry points are given in Fig. \ref{str}(c).
The horizontal dashed line indicates $E_F$.
}
\label{pmband}
\end{figure}

In the V $d$-orbital region, the blowup band structure with the $a_{1g}$ fatband 
is given in the left panel of Fig. \ref{pmband}.
Interestingly, the partially filled $e_{g}^\prime$ orbital is relatively well localized with a width of 0.7 eV,
leading to a sharp peak at 0.1 eV in the DOS.
The bonding and antibonding $a_{1g}$ orbitals below and above the $e_{g}^\prime$ orbital
have  an approximately twice larger bandwidth than that of the $e_{g}^\prime$ orbital 
due to the strong $dd\sigma$ interaction.
This is illustrated in the orbital-projected DOSs of the singlet $a_{1g}$, doublet $e_{g}^\prime$, 
and doublet $e_{g}$ manifolds, in the right panel of Fig. \ref{pmband}.
 
As given in the Inset of Fig. \ref{pmdos}, the atom-projected DOSs of V1 and V2 already show
some distinction, leading to a sizable charge difference of 0.08$e$ between them 
obtained from the Bader charge decomposition.
This value is almost independent of the magnetic states studied here.
This is consistent with the experimental observations,\cite{attfield,j.xing} 
in which the pure tetragonal-to-monoclinic structure transition leads to the CO.

\begin{figure}[tbp]
%\vskip 8mm
{\resizebox{8cm}{8cm}{\includegraphics{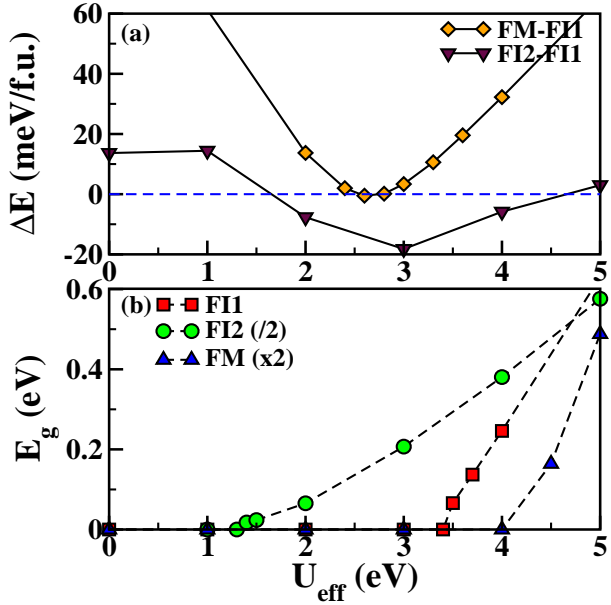}}}
\caption{Dependences on the strength of $U_{eff}$. 
(a) Variations in the energy differences $\Delta E$, with respect to the energy of FI1, 
 in units of meV per f.u.
 A positive value indicates that FI1 is energetically favored over the other state.
(b) Energy gaps $E_g$ of FM, FI1, and FI2.
 For a better visualization, $E_g$ of FM is enlarged by a factor of two, 
 while that of FI2 is divided by two.
}
\label{gap}
\end{figure}

\section{Ferrimagnetic state}
We have considered a few possible spin-ordered states including two different FI orders.
The FI1 state has the antiferrimagnetic intrachain and ferromagnetic interchain configuration,
whereas the FI2 state involves both intra- and inter-chain antiferrimagnetic interactions.
So in the FI2 state, the spins of the NN same-type V ions in the interlayer chain 
along the $\hat{b}$ direction, which is vertical to the ferrimagnetic V1-V2 chain, are also antialigned.
The magnitudes of the V local moments in FI1 and FI2 are almost identical,
but the net moment of FI2 is precisely compensated.  
In the GGA, as shown in Fig. \ref{gap}(a),
FI1 is energetically favored over the nonmagnetic (by 1.5 eV/f.u.),  FM (by 0.1 eV/f.u),
and FI2 (by 12 meV/f.u.) states, consistent with the experimental suggestions.\cite{attfield,j.xing}
Such a large magnetization energy is induced by the large spin moments of the V ions of 1.5--2.3 $\mu_B$.
For comparison, the energy gain due to a simple Stoner instability $IM^2/4$ leads to a similar value,
where the Stoner parameter $I\lesssim1$eV can be roughly estimated 
by the exchange splitting $\Delta_{ex}\approx IM\approx2.3$ eV (see below). 
Remarkably, the energetic preference among the spin-ordered states substantially depends on strength of $U_{eff}$,
suggesting a delicate magnetic character.
This issue will be further considered in the Discussion section.
In this section, we will focus on the FI1 state, which has been suggested from the experiments.\cite{attfield,j.xing}

\begin{figure}[tbp]
%\vskip 8mm
{\resizebox{8cm}{6.5cm}{\includegraphics{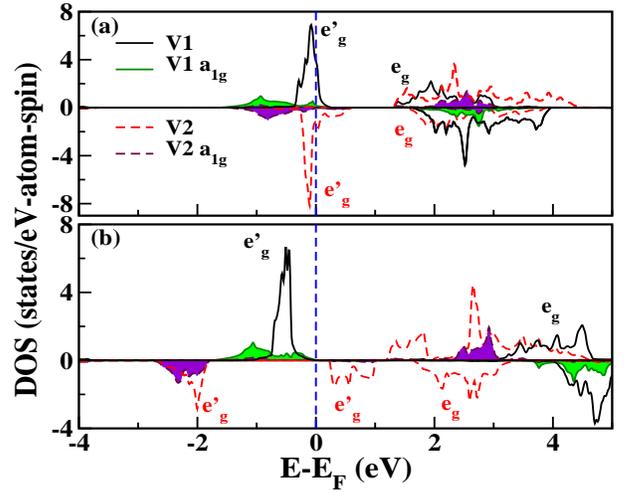}}}
\caption{FI1 orbital-projected DOSs of the V1 and V2 ions 
 obtained with the (a) GGA and (b) GGA+U at $U_{eff}=4$ eV.
 The orbitals of the V1 (V2) ion are represented by the solid (dashed) curves.
 The green (purple) shaded region indicates the V1 (V2) $a_{1g}$ orbital character.
 In the GGA, both the exchange and $t_{2g}-e_{g}$ crystal-field splittings are about 2.3 eV.
}
\label{fidos}
\end{figure}

\subsection{Uncorrelated regime: GGA level}
In the FI1 state, the V local moments are 1.94 (V1) and --1.71 (V2) $\mu_B$, 
leading to a net moment of 0.64 $\mu_B$/f.u.
The net moment is close to the value obtained in the polycrystalline sample,\cite{attfield}
but about twice that of the single crystalline sample.\cite{j.xing}
Considering the interstitial moment, these values are consistent with that of V$^{2.5+}$,
which is reflected in the DOS, presented in Fig. \ref{fidos}(a).

Figure \ref{fidos}(a) shows the orbital-projected DOSs of the V1 and V2 ions in the FI1 state.
$E_F$ is just above the peaks at --50 meV (--100 meV) for the up (down) channel.
The exchange splitting of the $t_{2g}$ manifold is about 2.3 eV in both V ions,
while the hybridization gap between the O $p$ orbital and this manifold is 2.5 eV,
yielding the well-isolated $t_{2g}$ manifolds with a bandwidth of 2.5 eV around $E_F$.
The partially filled $t_{2g}$ manifolds in the up and down channels
%roughly $\frac{5}{6}$-filled, 
are the V1 and V2 character, respectively.
In each majority $t_{2g}$ manifold, 
the centers of the completely filled $a_{1g}$ orbitals are approximately 1 eV lower than
those of the $e_g^\prime$ orbitals.
Besides, the V2 $e_g^\prime$ orbital in the down channel has 
a center higher by 50 meV and significantly longer unfilled tails than those of V1, 
which is in the up channel and is almost filled.
This feature leads to a sizable charge difference of 0.08$e$ between the V1 and V2 ions.
Additionally, upon the application of $U$ to the V ions, 
the almost filled V1 $e_g^\prime$ manifold is simply pushed down,
and a Mott transition occurs in the effectively half-filled $e_g^\prime$ manifold of V2 in the down channel (see below).

\begin{figure}[tbp]
%\vskip 8mm
{\resizebox{8cm}{6cm}{\includegraphics{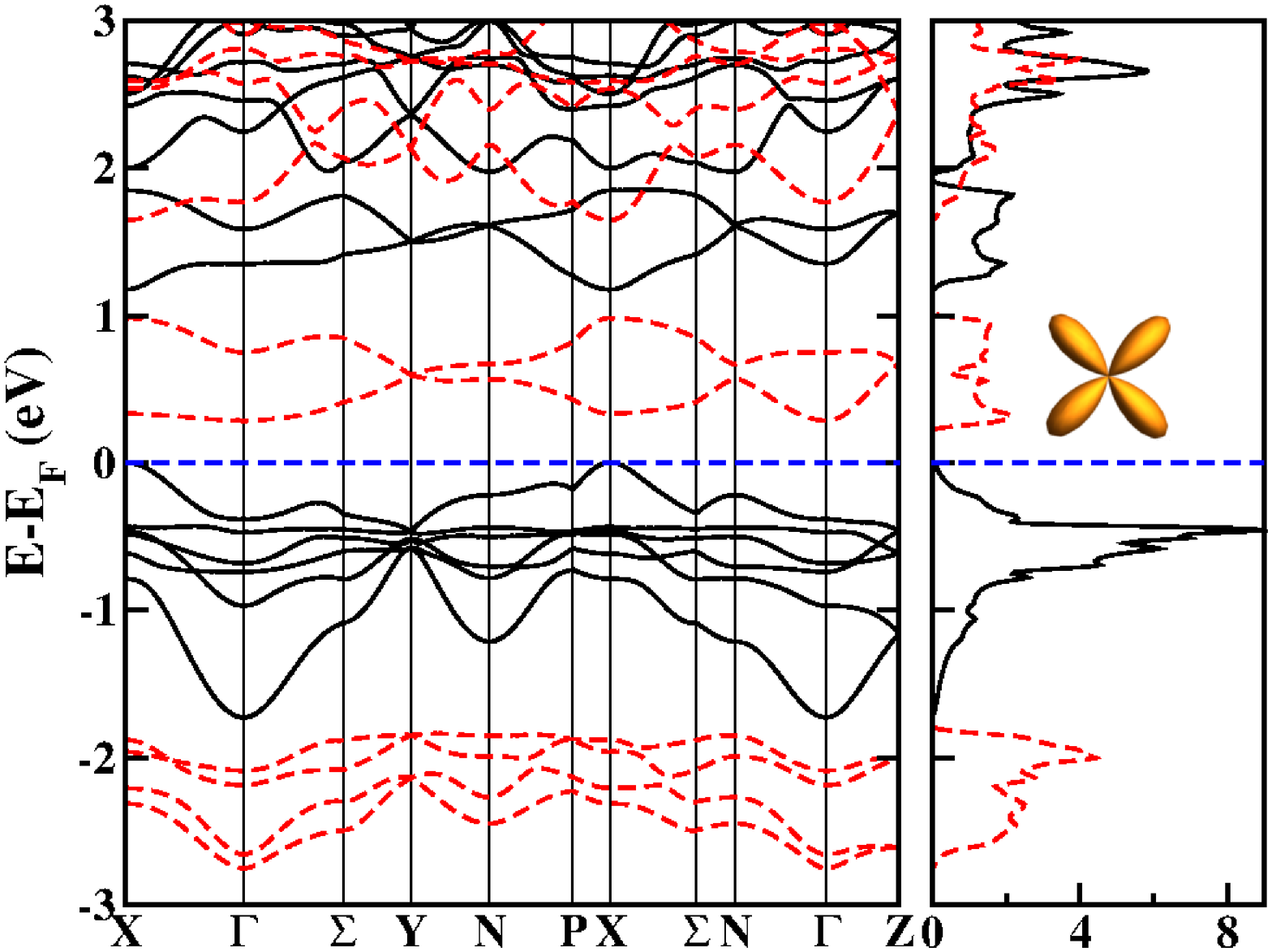}}}
\caption{ Left: FI1 band structure, in the region including only V $d$-orbitals, 
 obtained with GGA+U at $U_{eff}=4$ eV, where a spin-asymmetric gap of 0.25 eV is observed.
Right: Corresponding spin-resolved total DOSs in units of eV.
 The black-solid (red-dashed) curves represent the spin-up (-down) channel.
%The $a_{1g}$ orbital of V2 ion is highlighted by the (green) thick line.
Inset: Charge density isosurface of the upper Hubbard band, 
indicating the $e_{g,2}^{\prime}$ orbital.
}
\label{fiband_u}
\end{figure}

\subsection{Correlated regime: fully charge-ordered state}
The effects of correlations were considered with the GGA+U approach 
to obtain the observed insulating phase.
Figure \ref{gap}(b) shows a plot of $U_{eff}$ vs. energy gap $E_g$.
A gap opens at $U_{eff}^c\approx3.5$ (4.0) eV for the FI1 (FM) state,
similar to that of VO$_2$.\cite{vo2_dmft}
Interestingly, with the increase in $U_{eff}$ in the FM state, 
a transition from metal to half-semimetal with linearly crossing bands near $E_F$ (not shown here)
occurs at $U_{eff}\approx1$ eV, and then the energy gap is completely open at $U_{eff}\approx4$ eV.
Therefore, in the range of $U_{eff}\approx1-4$ eV, a magnetic Weyl semimetallic phase appears,
as suggested for the presumed tetragonal FM \vop.\cite{y.jin} 
On the other hand, in the FI2 state, $U_{eff}\approx1.5$ eV is sufficient to open a gap,
similar to the value of the antiferromagnetic V$_2$O$_3$.\cite{yu99}

%In V$_2$O$_3$, inclusion of correlation results in increasing a gap 
%between the filled $e_g^\prime$ and unfilled $a_{1g}$ manifolds, 
%which  are already sizably splitted by a trigonal distortion. 

Further, we focus on the FI1 insulating state at $U_{eff}=4$ eV, where a gap is clearly visible.
The band structure and total DOS in the region of the V $d$-orbitals are given in Fig. \ref{fiband_u}.
The corresponding orbital-projected DOSs are shown in Fig. \ref{fidos}(b).
In the down channel, the V2 $a_{1g}$ and one of the V2 $e_g^\prime$ orbitals around --2 eV are completely filled,
whereas the V1 $t_{2g}$ manifold is fully filled in the up channel.
The upper Hubbard band, {\it i.e.}, the other unfilled $e_g^\prime$ orbital, below 1 eV,
has the $e_{g,2}^\prime$ character, as shown in the Inset of Fig. \ref{fiband_u}.
Thus, the application of $U$ leads to an orbital ordering of $a_{1g}^{1\downarrow}e_g^{\prime{1\downarrow}}$ 
in V2 ($3d^2$), in contrast to the $e_g^{\prime{2}}$ configuration in the antiferromagnetic V$_2$O$_3$.
This leads to a spin-asymmetric gap of 0.25 eV between the spin-up V1 $t_{2g}$ and -down V2 $e_{g}^\prime$ bands.
%The gap considerably depends on spin channel, 1 eV in the up and 2 eV in the down.
In the state, the local moments are 2.30 $\mu_B$ (V1) and --1.54 $\mu_B$ (V2),
consistent with $S=3/2$ and 1, respectively.
Considering the interstitial moment and small contributions from the other ions, 
the total spin moment is 1 $\mu_B$/f.u.,
which is about three times larger than the experimentally observed value in the single crystalline sample.\cite{j.xing}
This issue is further discussed below.

\begin{figure}[tbp]
%\vskip 8mm
{\resizebox{7.0cm}{7.3cm}{\includegraphics{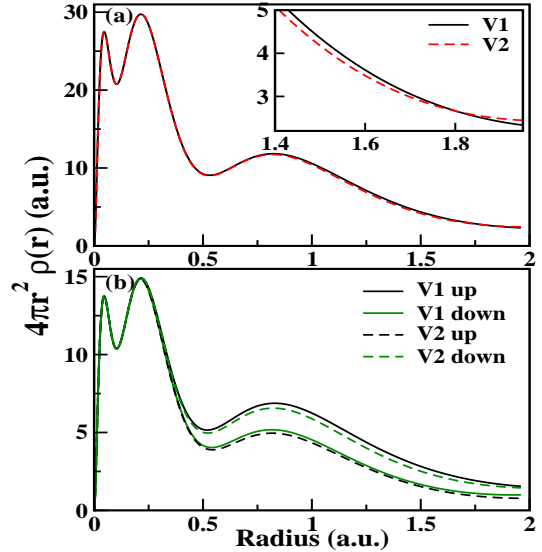}}}
\caption{(a) Radial charge densities $4\pi r^2\rho(r)$ of the V1 and V2 ions from each nucleus.
 The Inset represents an enlarged plot showing distinctions between the two ions.
(b) Spin-resolved charge densities of the V1 and V2 ions.
 Both are calculated in FI1 at $U_{eff}=4$ eV.
}
\label{radial}
\end{figure}

The CO can be analyzed with a few different approaches.\cite{ucd14}
Our calculated Bader charges $Q_B$ are $Q_B$(V1)=+1.58, $Q_B$(V2)=+1.94, $Q_B$(PO$_4$)=--2.23,
and $Q_B$(O3)=--1.30, 
leading to a large charge difference of 0.36$e$ between the V ions.
Compared with the formal values of +2, +3, --3, and --2, respectively, in the CO state,
this charge difference seems to be significantly smaller. 
However, this value is significantly larger than the value of 0.2$e$ 
observed in the charge disproportionated Na$_{1/2}$Co$_2$
with Co$^{3+}$ and Co$^{4+}$ ions.\cite{kwl05}
The summation of the occupation matrix gives 2.64 (2.38) for V1 (V2),
indicating the charge difference of 0.26$e$ between the two V ions,
consistent with the Bader charges.
Another approach is to analyze the radial charge densities $4\pi r^2\rho(r)$, 
as suggested by Quan, Pardo, and Pickett,\cite{ucd14,ucd12} 
which have a clear difference in their tails in a CO state.
Figure \ref{radial}(a) shows the densities of the two V ions, obtained from {\sc wien2k}.
Below approximately 0.5 a.u., the charge densities insignificantly differ,
while a difference is observed above this value.
This distinction is more clearly observed in the spin-resolved charge densities in Fig. \ref{radial}(b),
particularly in the majority densities.

\begin{figure}[tbp]
%\vskip 8mm
{\resizebox{8cm}{6cm}{\includegraphics{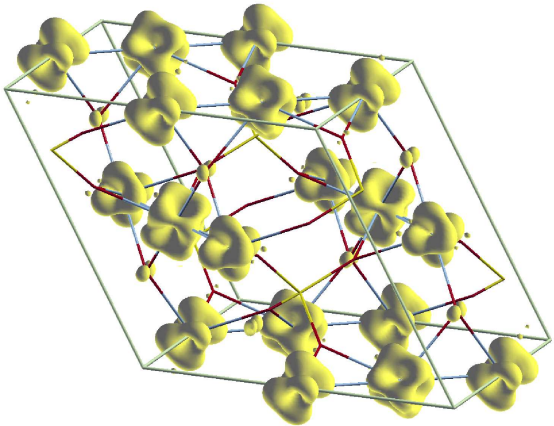}}}
\caption{Charge density plot with the isosurface at 0.025 $e$/\AA$^3$~
in the insulating FI1, showing the orbital order.
These shapes indicate an equally occupied $t_{2g}$ orbital for V1 and
$a_{1g}$ plus $e_{g,1}^\prime$orbitals for V2.
The O3 ions show an obvious $p_\pi$ character,
indicating a sizable $pd\pi$ interaction along the vertical direction.
The densities of the other O ions are invisible.
}
\label{char}
\end{figure}

Figure \ref{char} shows the charge density isosurface in GGA+U at $U_{eff}=4$ eV.
The V1 ion has the shape of the equally occupied $t_{2g}$ orbital, 
{\it i.e.}, $t_{2g}^{3\uparrow}$.
On the other hand, the shape of the V2 ion indicates  a combination of the $a_{1g}$ and $e_{g,1}^\prime$ orbitals.
So, the lower Hubbard band at --2 eV in the spin down is the $e_{g,1}^\prime$ orbital.
Consequently, the V2 ions show the $a_{1g}^{1\downarrow}e_{g,1}^{\prime1\downarrow}$ orbital order.
%The orbital order in V2 ions is different from the antiferromagnetic V$_2$O$_3$ of  the $e_{g}^{\prime2\uparrow}$.
However, this order leads to only a small orbital moment, 
since the occupation matrix of V2 is nearly symmetric.
Our GGA+U+spin-orbit coupling (SOC) calculations show that the orbital moment is about 0.04 $\mu_B$ in each V ion,
opposite to each spin moment.
This indicates that this order cannot be attributed to the significantly reduced net moment 
in the experiments.\cite{attfield,j.xing}
It is worth noting that the O3 ion has a small moment, but exhibits a clear $p_\pi$ character. 
This implies a considerable $pd\pi$ interaction,
which would lead to the next nearest neighbor (NNN) superexchange pattern 
with an angle of 130$^\circ$ along the $\hat{b}$ direction.

\section{Discussion on quantum fluctuation}
As mentioned above,
the experimentally observed moments in the system are approximately 1/3 of our calculated values.
These discrepancies are significant,
although the theoretically calculated moment should not be expected to be identical 
to the experimentally observed value, considering the theoretical specifications
(and some spin in the interstitial region) in the theoretical approach. 
Below, we discuss a possible mechanism of such a large reduction.

\subsection{Experimental indications}
There are several experimental indications of complex unusual magnetic properties
of this system.\cite{j.xing}
The magnetization exhibits a small magnetic anisotropy,
and the magnetic entropy is considerably smaller than those of V$^{2+}$ and V$^{3+}$.
The observed Curie-Weiss and ordering temperatures lead to 
a sizable frustration parameter of $|\theta_{CW}|/T_C\approx 7.03$,
comparable with the typical values of frustrated systems.\cite{frust}
As mentioned above, the V ions form a pyrochlore-like link consisting of
the V$_4$ tetrahedra with two sides of 2.67 \AA~ and four sides of $\sim$3.62 \AA.
The long sides are longer by 1/3 than the short sides, but the length is relatively short,
compared with those of the existing candidates of 1D chain and frustrated structure.
For example, the NNN distance is 4.40 -- 4.92 \AA~ 
in $\beta-$TeVO$_4$ of the NN 3.64 \AA,\cite{tvo11}.
In Ca$_3$Co$_2$O$_6$ and Sr$_2$Rh$_4$O$_{12}$ 
the NNN distances are twice longer than the NN distances of $\sim$2.6 \AA.\cite{hardy04,parkin07}
Additionally, this system shows a small coercivity of the order of mT,\cite{j.xing} 
which implies an easily variable magnetization direction by an external magnetic field or magnetic impurity.
It may be expected that the complex  behavior originates from strong quantum fluctuations or frustration.

\subsection{Evaluation of the magnetic ground state}
We considered three magnetic states, FM, FI1, and FI2.
Figure \ref{gap}(a) shows the evaluation of differences in energy between these three states,
as varying the strength of $U_{eff}$.
Remarkably, with the increase in $U_{eff}$,
the difference in energy between FM and FI1 is reduced and becomes negative around $U_{eff}\approx2.6$ eV,
indicating that FM is favored over FI1 in this regime.
Above this value, the energy difference increases again, leading to an energetically favored FI1 state.
Comparing FI1 with FI2, the FI2 state is energetically favored over the FI1 state 
in the range of $U_{eff}$=1.5 to 4.7 eV.
Even at $U_{eff}=5$ eV, the energy difference between FI1 and FI2 is just a few meV.
This indicates that the NNN superexchange interaction is also crucial in this system. 
Since this behavior is observed in the reasonable range of $U_{eff}$ for this vanadium oxide,
our results suggest that frustration would be a reliable scenario in this system.

Additionally, as expected from the tiny orbital moment,
this system shows a little magnetic anisotropy, obtained from GGA+SOC calculations.
We considered two quantized directions parallel and perpendicular to the chain.
The anisotropy energy is less than a few tenths meV/f.u., 
which indicates that the application of a small magnetic field
can vary the spin direction.
This also supports the proposed scenario.

\begin{figure}[tbp]
%\vskip 8mm
{\resizebox{8cm}{6cm}{\includegraphics{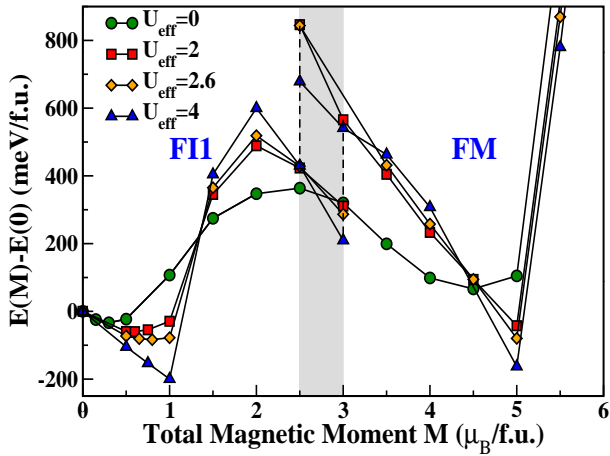}}}
\caption{Energy versus fixed spin moment $M$ plot,
 for several values of $U_{eff}$ (in eV).
 At the shaded region, both FI1 and FM states coexist in the GGA+U cases.
Below this region, the FI1 state appears, whereas the FM state is observed above the region.
 Note that the antialigned moments of V1 and V2 are canceled by each other at $M=0$
 in the FI1 state, whose energy is higher than that of FI2. 
}
\label{fsm}
\end{figure}

\subsection{Fixed spin moment studies}
In order to affirm the delicate magnetic behavior, fixed spin moment (FSM) calculations were performed,\cite{fsm}
as varying the strength of $U_{eff}$.
In these calculations, two states, FI1 and FM, were considered by choosing a proper initial condition.
Figure \ref{fsm} shows the energy difference $\Delta E(M)$ plotted against the total moment $M$, 
with respect to the energy at $M=0$ for each $U_{eff}$, 
where the net moment is completely canceled by the antialigned V1 and V2.
These results are consistent with our self-consistent results, given in Fig. \ref{gap}(a).

Interestingly, two minima appear, regardless of the value of $U_{eff}$.
One of them is in the range of 0.5 -- 1 $\mu_B$ for the FI1 state, 
while the other is in the range of 4.5 -- 5 $\mu_B$ for the FM state.
In the GGA ($U_{eff}=0$),
a large range of magnetic field $\Delta B=\Delta E/\Delta M$ 
is required for the magnetic phase transition from FI1 to FM.
However, as increasing $U_{eff}$,
the energy difference $\Delta E$ between the two states decreases 
and becomes close to zero around $U_{eff}=2.5$ eV.
So, only a small magnetic field is required for the phase transition around this value of $U_{eff}$.
Above this value, $\Delta E$ monotonously increases reaching 37 meV at $U_{eff}=4$ eV, 
where both states are insulating.
%suggesting a magnetic field $\Delta B$=160 T for the phase transition.
Thus, these results also suggest that this system is close to a quantum fluctuation.

Note that the $\Delta E(M)$ curves are discontinuous and  
both FI1 and FM states appear coincidentally in the range of $M=2.5$ -- 3 $\mu_B$ in the GGA+U cases,
leading to hysteresis. 
This has been observed in the density functional theory plus $U$ approach, which often leads to multisolutions
in a critical region.\cite{kwl05,kwl06}

%\begin{eqnarray}
%\label{eqn}
%\end{eqnarray}

\section{Summary}
Through {\it ab initio} calculations including correlation effects, 
we have investigated the insulating $\beta$-\vop~ with a quasi-1D ferrimagnetic Heisenberg spin chain structure,
exhibiting the charge and spin orderings
at $T_{CO}\approx600$ K and $T_{SO}\approx125$ K, respectively.
Remarkably,this system has a weak pyrochlore-like tetrahedral link of V ions 
due to the relatively short interlayer (NNN) distance along the vertical direction.

Consistent with the experimental observation,\cite{attfield}
our calculations indicate that the charge order of alternating V$^{3+}$ and $V^{2+}$ along the chain
occurs coincidentally with the structure transition from tetragonal to monoclinic.
In the experimentally suggested FI1 state, 
where the spins of the NN V ions are antialigned along the chain,
inclusion of correlation effects leads to a full orbital polarization of 
V1 ($t_{2g}^{3\uparrow}$, $S=\frac{3}{2}$) 
and V2 ($a_{1g}^{1\downarrow}e_g^{\prime{1\downarrow}}$, $S=1$),
above the critical value of $U_{eff}^c=$3.5 eV for the metal-insulator transition.
This yields the net moment of about 1 $\mu_B$ (including the small orbital moments) 
which is 2--3 times larger than the experimentally observed value.
Besides, the energy differences between the few spin-ordered states show 
the strong $U$-dependence, which is also supported by the FSM calculations.
This implies that a magnetic phase transition is feasible and 
that the NNN superexchange interaction through the $pd\pi$ hybridization is sizable.
In addition to the several experimental indications mentioned above,
the significant variations in energy differences, and our calculated tiny orbital moments 
of several hundredth $\mu_B$,
the substantially quenched moment suggests that quantum fluctuation (frustration)
occurs in the pyrochlore-like tetrahedral structure.
 
Our results suggest that this system is a promising candidate to show the vital interplay 
among the charge-, spin-, and lattice-degrees of freedom, and geometrical frustration. 
Further experimental and theoretical researches are required  
to elucidate the complex magnetic properties.

\section{Acknowledgments}
We acknowledge K.-H. Ahn and Y.-J. Song for useful discussions on calculation methods.
This research was supported by NRF of Korea Grant No. NRF-2016R1A2B4009579.


\begin{thebibliography}{10}
%low-dimension
\bibitem{gros03} For a review, see P. Lemmens, G. G\"untherodt, and C. Gros,
Magnetic light scattering in low-dimensional quantum spin systems,
Phys. Rep. {\bf 375}, 1 (2003).

\bibitem{wang15} Z. Wang, M. Schmidt, A. K. Bera, A. T. M. N. Islam, B. Lake, A. Loidl,
and J. Deisenhofer,
Spinon confinement in the one-dimensional Ising-like antiferromagnet SrCo$_2$V$_2$O$_8$,
 Phys. Rev. B {\bf 91}, 140404(R) (2015).

\bibitem{rice96} E. Dagotto and T. M. Rice,
 Surprises on the way from one- to two-dimensional quantum magnets: the ladder materials,
 Science {\bf 271}, 618 (1996).


%low-dimensional spin chain & frustration
\bibitem{hida08} K. Hida and K. Takano,
 Frustration-induced quantum phases in mixed spin chain with frustrated side chains,
 Phys. Rev. B {\bf 78}, 064407 (2008).

\bibitem{star15} For a recent review, see O. A. Starykh,
 Unusual ordered phases of highly frustrated magnets: a review,
 Rep. Prog. Phys. {\bf 78}, 052502 (2015).


%lowD+frust. (triangle)
\bibitem{hardy03} V. Hardy, S. Lambert, M. R. Lees, and D. Mck. Paul,
Specific heat and magnetization study on single crystals of the frustrated quasi-one-dimensional
oxide Ca$_3$Co$_2$O$_6$,
 Phys. Rev. B {\bf 68}, 014424 (2003).

\bibitem{hardy04} V. Hardy, M. R. Lees, O. A. Petrenko, D. Mck. Paul,
D. Flahaut, S. H\'ebert, and A. Maignan,
 Temperature and time dependence of the field-driven magnetization steps
 in Ca$_3$Co$_2$O$_6$ single crystals,
 Phys. Rev. B {\bf 70}, 064424 (2004).

\bibitem{tvo11} Yu. Savina, O. Bludov, V. Pashchenko, S. L. Gnatchenko, P. Lemmens, and H. Berger,
 Magnetic properties of the antiferromagnetic spin-$\frac{1}{2}$ chain system $\beta-$TeVO$_4$,
 Phys. Rev. B {\bf 84}, 104447 (2011).

\bibitem{tvo16} F. Weickert, N. Harrison, B. L. Scott, M. Jaime, A. Leitm\"ae, I. Heinmaa,
 R. Stern, O. Janson, H. Berger, H. Rosner, and A. A. Tsirlin,
 Magnetic anisotropy in the frustrated spin-chain compound $\beta-$TeVO$_4$,
 Phys. Rev. B {\bf 94}, 064403 (2016).

\bibitem{parkin07} G. Cao, V. Durairaj, S. Chikara, S. Parkin, and P. Schlottmann,
 Partial antiferromagnetism in spin-chain Sr$_5$Rh$_4$O$_{12}$, Ca$_5$Ir$_3$O$_{12}$, 
 and Ca$_4$IrO$_6$ single crystals,
 Phys. Rev. B {\bf 75}, 134402 (2007).

\bibitem{ladder} V. Ravi Chandra, N. B. Ivanov, and J. Richter,
 Frustrated spin ladder with alternating spin-1 and spin-$\frac{1}{2}$ rungs,
 Phys. Rev. B {\bf 81}, 024409 (2010).


%Vanadium: V2O3(d2), VO2(d1)
\bibitem{cryfield} A. Tanaka,
 Electronic structure and phase transition in V$_2$O$_3$:
 Importance of $3d$ spin-orbit interaction and lattice distortion,
 J. Phys. Soc. Jpn. {\bf 71}, 1091 (2002).

\bibitem{v2o3_08} M. M. Qazilbash, A. A. Schafgans, K. S. Burch, S. J. Yun, B. G. Chae,
 B. J. Kim, H. T. Kim, and D. N. Basov,
 Electrodynamics of the vanadium oxides VO$_2$ and V$_2$O$_3$,
 Phys. Rev. B {\bf 77}, 115121 (2008).

\bibitem{v2o3_06} M. S. Laad, L. Craco, and E. M\"{u}ller-Hartmann,
 Orbital-selective insulator-metal transition in V$_2$O$_3$ under external pressure,
 Phys. Rev. B {\bf 73}, 045109 (2006).

\bibitem{yu99} S. Yu. Ezhov, V. I. Anisimov, D. I. Khomskii, and G. A. Sawatzky,
 Orbital Occupation, Local Spin, and Exchange Interactions in V$_2$O$_3$,
 Phys. Rev. Lett. {\bf 83}, 4136 (1999).

\bibitem{vo2_dmft} S. Biermann, A. Poteryaev, A. I. Lichtenstein, and A. Georges,
 Dynamical Singlets and Correlation-Assisted Peierls Transition in VO$_2$,
 Phys. Rev. Lett. {\bf 94}, 026404 (2005).

\bibitem{v2o3_18} F. Lechermann, N. Bernstein, I. I. Mazin, and R. Valent\'i,
Uncovering the Mechanism of the Impurity-Selective Mott Transition in Paramagnetic V$_2$O$_3$,
 Phys. Rev. Lett. {\bf 121}, 106401 (2018).

\bibitem{mrs17} M. Brahlek, L. Zhang, J. Lapano, H.-T. Zhang, R. Engel-Herbert,
 N. Shukla, S. Datta, H. Paik, and D. G. Schlom, 
 Opportunities in vanadium-based strongly correlated electron systems,
 MRS Commun. {\bf 7}, 27 (2017).

\bibitem{botana11} A. S. Botana, V. Pardo, D. Baldomir, A. V. Ushakov, and D. I. Khomskii,
 Electronic structure of V$_4$O$_7$: Charge ordering, metal-insulator transition, and magnetism,
 Phys. Rev. B {\bf 84}, 115138 (2011).


%exp
\bibitem{glaum} R. Glaum and R. Gruehn,
 Synthese, Kristallstruktur und magnetisches Verhalten von $\beta$-V$_2$PO$_5$,
 Z. Kristallogr. {\bf 186}, 91 (1989).

\bibitem{y.jin} Y. J. Jin, R. Wang, Z. J. Chen, J. Z. Zhao, Y. J. Zhao, and H. Xu,
 Ferromagnetic Weyl semimetal phase in a tetragonal structure,
 Phys. Rev. B {\bf 96}, 201102(R) (2017).
 
\bibitem{attfield} E. Pachoud, J. Cumby, C. T. Lithgow, and J. P. Attfield, 
 Charge Order and Negative Thermal Expansion in V$_2$OPO$_4$,
 J. Am. Chem. Soc. {\bf 140}, 636 (2018).

\bibitem{j.xing} J. Xing, H. Cao, A. Paul, C. Hu, H.-H. Wang, Y. Luo, R. Chaklashiya, 
 J. M. Allred, S. Brown, T. Birol, and N. Ni,
 Charge ordering and ferrimagnetism in the strongly correlated $\beta$-V$_2$PO$_5$ single crystal,
 arXiv:1712.09973 (2017).


%methods
\bibitem{gga} J. P. Perdew, K. Burke, and M. Ernzerhof,
 Generalized gradient approximation made simple,
 Phys. Rev. Lett. {\bf 77}, 3865 (1996).

\bibitem{wien2k} K. Schwarz and P. Blaha,
 Solid state calculations using WIEN2k,
 Comput. Mater. Sci. {\bf 28}, 259 (2003).

\bibitem{amf} V. I. Anisimov, I. V. Solovyev, M. A. Korotin, M. T. Czyzyk, and G. A. Sawatzky, 
 Density-functional theory and NiO photoemission spectra,
 Phys. Rev. B {\bf 48}, 16929 (1993).

%\bibitem{erik09} E. R. Ylvisaker, K. Koepernik, and W. E. Pickett,
%  Anisotropy and Magnetism in the LSDA+U Method,
% Phys. Rev. B {\bf 79}, 035103 (2009).

%CD
\bibitem{kwl05} K.-W. Lee, J. Kun\v{e}s, P. Novak, and W. E. Pickett,
 Disproportionation, Metal-Insulator Transition, and Critical Interaction Strength in Na$_{1/2}$CoO$_2$,
 Phys. Rev. Lett. {\bf 94}, 026403 (2005).


%Radial charge
\bibitem{ucd14} W. E. Pickett, Y. Quan, and V. Pardo,
 Charge states of ions, and mechanisms of charge ordering transitions,
 J. Phys.: Condens. Matter {\bf 26}, 274203 (2014).

\bibitem{ucd12} Y. Quan, V. Pardo, and W. E. Pickett,
 Formal Valence, 3d-electron Occupation, and Charge-Order Transitions,
 Phys. Rev. Lett. {\bf 109}, 216401 (2012).


%frust. parameter
\bibitem{frust} A. P. Ramirez,
 Strongly Geometrically Frustrated Magnets,
 Annu. Rev. Mater. Sci. {\bf 24}, 453 (1994).

%FSM
\bibitem{fsm} K. Schwarz and P. Mohn, 
 Itinerant metamagnetism in YCo$_2$,
J. Phys. F:Met. Phys. {\bf 14}, L129 (1984).

%fsm in U
\bibitem{kwl06} K.-W. Lee and W. E. Pickett,
 Correlation effects in the high formal oxidation-state compound Sr$_2$CoO$_4$,
 Phys. Rev. B {\bf 73}, 174428 (2006).

\end{thebibliography}
\end{document}